\documentclass{article}

\usepackage{PRIMEarxiv}

\usepackage[utf8]{inputenc} % allow utf-8 input
\usepackage[T1]{fontenc}    % use 8-bit T1 fonts
\usepackage{url}            % simple URL typesetting
\usepackage{booktabs}       % professional-quality tables
\usepackage{amsfonts}       % blackboard math symbols
\usepackage{nicefrac}       % compact symbols for 1/2, etc.
\usepackage{microtype}      % microtypography
\usepackage{lipsum}
\usepackage{fancyhdr}       % header
\usepackage{graphicx}       % graphics
\usepackage{multirow}
\usepackage{amsmath,amssymb}
\usepackage{tabularx}       % for flexible column widths
\usepackage{adjustbox}      % for scaling tables
\usepackage{rotating}       % for sideways tables
\usepackage{array}          % for better column formatting
\graphicspath{{media/}}     % organize your images and other figures under media/ folder

\usepackage[
    colorlinks=true,        % color links instead of boxes
    linkcolor=blue,         % color of internal links
    citecolor=blue,         % color of citations
    urlcolor=blue,          % color of URLs
    filecolor=blue,         % color of file links
    bookmarks=true,         % show bookmarks bar
    breaklinks=true         % allow links to break over lines
]{hyperref}

\title{The Consistency-Acceptability Divergence of LLMs in Judicial Decision-Making: Task and Stakeholder Dimensions
\thanks{\textit{\underline{Citation}}: 
\textbf{Zhang, M., Xu, Q. The Consistency-Acceptability Divergence of LLMs in Judicial Decision-Making: Task and Stakeholder Dimensions. arXiv:XXXX.XXXXX [cs.CY]}} 
}

\author{
  Mingda Zhang \\
  School of Software \\
  Yunnan University \\
  Yuhua Road, Kunming 650500, Yunnan, China\\
  \texttt{mingda.zhang@ynu.edu.cn} \\
   \And
  Qing Xu \\
  School of Law \\
  Yunnan University \\
  Cuihu North Road, Kunming 650091, Yunnan, China\\
  \texttt{20180090@ynu.edu.cn} \\
}

\begin{document}
\maketitle

\begin{abstract}
The integration of large language model (LLM) technology into judicial systems is fundamentally transforming legal practice worldwide. However, this global transformation has revealed an urgent paradox requiring immediate attention. This study introduces the concept of ``consistency-acceptability divergence'' for the first time, referring to the gap between technical consistency and social acceptance. While LLMs achieve high consistency at the technical level, this consistency demonstrates both positive and negative effects. Through comprehensive analysis of recent data on LLM judicial applications from 2023-2025, this study finds that addressing this challenge requires understanding both task and stakeholder dimensions. This study proposes the Dual-Track Deliberative Multi-Role LLM Judicial Governance Framework (DTDMR-LJGF), which enables intelligent task classification and meaningful interaction among diverse stakeholders. This framework offers both theoretical insights and practical guidance for building an LLM judicial ecosystem that balances technical efficiency with social legitimacy.
\end{abstract}

% keywords can be removed
\keywords{Large language models \and Judicial decision-making \and Consistency-acceptability divergence \and Instrumental and value rationality \and Technology acceptance \and AI governance}

\section{Introduction}

The integration of large language model (LLM) technology into judicial systems represents a fundamental shift in how legal practice operates globally. Since ChatGPT became publicly available, LLM applications in law have grown rapidly, expanding from basic document processing and legal research to core judicial decision-making activities \cite{ref1}. For instance, the Shenzhen Intermediate Court's integration of LLMs into its decision-making processes marks a significant milestone in practical implementation. This trend has spread to multiple jurisdictions, including the United States, United Kingdom, Colombia, and India, making it a truly global phenomenon \cite{ref2,ref3}.

However, LLMs demonstrate clear dual-edged characteristics when pursuing decision consistency. Recent findings from Apple's research team reveal an important insight: artificial intelligence (AI) ``reasoning'' models, including Claude, DeepSeek-R1, and o3-mini, do not actually perform genuine reasoning. Instead, they rely on pattern memorization \cite{ref4}. While the Shenzhen Court's implementation achieved notable efficiency gains, it also exposed a critical limitation---the mechanical consistency based on pattern memorization rather than true reasoning. Research shows that GPT-4o functioning as ``Judge AI'' behaves like a formalistic judge, heavily influenced by precedents but lacking empathy, unlike human judges. This results in typical ``machine coldness'' characteristics \cite{ref5}. Although this consistency built on pattern memorization appears impressive in technical metrics, it becomes a major barrier to social recognition and legitimacy.

This excessive focus on consistency has created an acceptability crisis in judicial applications. While LLMs achieve consistency through standardization and patterning, which enhances formal efficiency, this approach triggers widespread social acceptance barriers. The reason is that it neglects the contextual nature and value pluralism inherent in judicial decision-making. This consistency-acceptability divergence shows systematic characteristics across both task and stakeholder dimensions. In the task dimension, technical tasks gain higher acceptance because their clear rules and verifiable results align well with consistency logic. In contrast, value-judgment-intensive judicial tasks face sharp drops in acceptability due to mechanical consistency. This mechanical approach not only fails to handle case specificity but also produces automation bias---people need to know the correct answer beforehand to identify LLM errors. In the stakeholder dimension, different groups show varied attitudes based on their positions: judges view consistency as threatening judicial independence and show low acceptance; lawyers passively accept it under efficiency pressure while harboring internal doubts; the public experiences the contradiction of recognizing technological potential while lacking substantive trust; and marginalized groups oscillate between hoping for fairness and fearing bias perpetuation. These differentiated patterns reflect how consistency, as instrumental rationality, conflicts with the value rationality held by different groups \cite{ref6,ref7}.

Weber's rationalization theory provides a useful framework for understanding this phenomenon. He distinguished between instrumental rationality (pursuing maximum efficiency, standardization, and calculability) and value rationality (pursuing substantive justice based on beliefs). This distinction reveals the ``iron cage'' dilemma caused by expanding instrumental rationality in modernization \cite{ref8}. In LLM judicial applications, this tension appears particularly acute. On one hand, LLMs achieve judgment standardization through algorithmic consistency, representing the ultimate instrumental rationality---reducing complex legal reasoning to calculable pattern matching. On the other hand, the essence of justice requires value rationality---understanding case particularity, empathizing with parties' circumstances, and pursuing social justice. This structural conflict between instrumental and value rationality forms the deep root of the consistency-acceptability divergence.

Habermas's communicative rationality theory offers a potential solution for reconciling this tension. Through rational dialogue among different parties, it is possible to acknowledge instrumental efficiency while maintaining value pluralism. This allows technical logic and justice demands to achieve dynamic balance through equal deliberation \cite{ref9,ref10}. However, current practice lacks this communicative bridge. Technical systems optimize consistency metrics following instrumental rationality logic, while social concerns based on value rationality cannot be effectively integrated. Judges fear professional judgment replacement, lawyers sacrifice value considerations under efficiency pressure, and the public distrusts mechanized justice---all manifestations of this conflict. Without communicative rationality mechanisms, consistency's instrumental advantages cannot translate into social acceptance due to lacking value recognition. This ultimately becomes the core obstacle to establishing legitimacy for LLM judicial applications \cite{ref11}.

The absence of communicative rationality mechanisms not only intensifies the conflict between instrumental and value rationality but also fundamentally undermines the social foundation for LLM judicial applications. The core challenge is constructing a social legitimacy foundation---achieving widespread recognition of both technical legitimacy and legality at the social level. The European Union's classification of judicial AI as high-risk systems through the Artificial Intelligence Act, requiring strict data quality control and human supervision, reflects this concern \cite{ref12}. The lack of legitimacy recognition stems from multiple factors. Technically, LLMs' statistical learning architecture struggles to understand the deep logic and value connotations of legal reasoning. Institutionally, effective cross-group deliberation mechanisms are lacking. Culturally, different judicial traditions lead to varied responses. Only by recognizing how consistency's dual nature produces different effects across tasks and groups, and by constructing governance frameworks that leverage consistency's advantages while avoiding negative impacts, can LLM judicial applications obtain sustainable social foundations \cite{ref13}.

To address these theoretical gaps and practical challenges, this study aims to achieve three core objectives:

First, this study conceptualizes the core paradox in LLM judicial applications as the consistency-acceptability divergence. By integrating the latest empirical research, this study systematically analyzes how this divergence emerges, revealing how consistency undermines substantive justice while enhancing efficiency. This study examines the structural conflict between instrumental and value rationality and establishes a theoretical framework for evaluating this phenomenon.

Second, this study constructs a dual-dimensional analytical framework to reveal the divergence's essence. From the task dimension, this study analyzes differentiated acceptability effects and deep fractures across different judicial tasks. From the stakeholder dimension, this study examines differentiated response patterns and structural constraints among different groups. This approach helps deeply understand the social roots of the divergence phenomenon.

Third, this study constructs LLM judicial governance theory based on social legitimacy foundations. By integrating Weber's rationalization theory and Habermas's communicative rationality framework, this study proposes the Dual-Track Deliberative Multi-Role LLM Judicial Governance Framework (DTDMR-LJGF). This framework represents a paradigm shift from technological determinism to rational pluralistic integration.

\section{Results}

\subsection{Judicial Task Dimension: Cross-Task Value Rationality Differences and Technical Boundaries}

AI application acceptance in the judicial field shows significant variation across different tasks, clearly demonstrating consistency's dual-edged effects. To systematically analyze this phenomenon, this study compiled the latest empirical data from major legal industry research institutions and academic institutions between 2024-2025. Table~\ref{tab:task-analysis} synthesizes acceptance data covering nine judicial task types. Data sources include large-scale surveys from authoritative institutions such as the Clio Legal Trends Report, Thomson Reuters Future of Professionals Report, and Stanford RegLab, as well as practical data from U.S. judicial systems, corporate legal departments, and international law firms.

These data reflect a complete spectrum from highly standardized auxiliary tasks (such as legal research and contract review) to decision-making tasks requiring intensive value judgments (such as judicial decision support and sentencing recommendations). Through systematic compilation of acceptance indicators, rational characteristics, and data sources for each task type, Table~\ref{tab:task-analysis} demonstrates not only LLMs' differentiated adaptability across judicial tasks but also reveals how consistency produces dramatically different social effects in different contexts. This variation ultimately affects the establishment of legitimacy foundations.

\begin{table*}[bt!]
\centering
\small % Reduce font size
\caption{Differentiated Analysis of LLM Acceptance in Legal Tasks (2024-2025)}
\label{tab:task-analysis}
\begin{adjustbox}{width=\textwidth,center}
\begin{tabular}{p{2.8cm}p{3.8cm}p{3.8cm}p{3.2cm}p{1.4cm}}
\toprule
\textbf{Task Type} & \textbf{Research Content} & \textbf{Key Indicators} & \textbf{Rational Characteristics} & \textbf{Data Sources}\\
\midrule
Legal Research \& Case Analysis & Survey of 1,028 legal professionals analyzing AI adoption growth trends & 79\% of lawyers use LLM tools, 60\% year-over-year increase & Mature technology, controlled risks, significant efficiency gains & \cite{ref14,ref15}\\
\addlinespace[3pt]
Contract Review & Survey of 800 lawyers and 712 international legal professionals & 74\% use LLM tools, 92\% believe it improves work & High standardization, verifiable accuracy & \cite{ref16,ref17}\\
\addlinespace[3pt]
Document Review & Using fine-tuned Mistral 7B model to detect racial restrictive covenants & 99\% recall rate, 100\% precision, saving 86,500 person-hours & Cost less than 2\% compared to proprietary models & \cite{ref18}\\
\addlinespace[3pt]
E-Discovery & Survey of 186 organizations (65\% Fortune 500) evaluating Technology-Assisted Review & Only 12\% corporate use, but courts widely accept TAR & Mature technology, accuracy superior to manual review & \cite{ref19}\\
\addlinespace[3pt]
Transcription Services & Survey of legal professionals from various firm sizes & 22.1\% of law firms use & Mature technology, continuously improving accuracy & \cite{ref20}\\
\addlinespace[3pt]
Legal Consultation \& Advice & Survey of 1,200+ international professionals & 26\% actively use generative AI, but 83\% consider AI providing legal advice inappropriate & Involves professional core values, higher ethical risks & \cite{ref7,ref21}\\
\addlinespace[3pt]
Court Representation & Research on AI application possibilities in court representation & 96\% believe AI representing clients in court "goes too far" & Interpersonal trust irreplaceable, legal reasoning complexity & \cite{ref7}\\
\addlinespace[3pt]
Judicial Decision Support & Survey of 1,000 adults; Fine et al. experimental study with 1,800 participants & Judicial system trust drops to 35\%, pure expertise judges significantly preferred & Systematic risk of algorithmic bias, threat to procedural justice & \cite{ref22,ref6}\\
\addlinespace[3pt]
Sentencing Recommendations & Evaluation of three major legal AI research tools reliability & Professional legal LLM tools show 17\%+ hallucination rate (Westlaw AI exceeds 34\%) & Prominent bias and errors, fairness controversies & \cite{ref23}\\
\bottomrule
\end{tabular}
\end{adjustbox}
\end{table*}

Based on the analysis of acceptance differentiation across judicial tasks, this study identifies three essential mechanisms. These mechanisms collectively reveal how consistency's dual nature produces differentiated effects across task types, thereby affecting the social legitimacy foundation of LLM judicial applications.

First, this study observes an epistemological fracture: the fundamental incompatibility between knowledge representation and meaning generation. The surface-level task acceptance gradient---from high acceptance of technical tasks to low acceptance of creative tasks---essentially reflects a divide between two fundamentally different ways of understanding knowledge. Technical tasks achieve high acceptance rates, such as legal research (79\% lawyer usage), contract review (74\% usage rate, 92\% believe it improves work), and document processing (99\% recall rate, 100\% precision). These tasks succeed because they essentially involve knowledge representation---encoding, retrieving, and recombining established information according to formal rules \cite{ref24}. In these contexts, consistency becomes a positive force for enhancing efficiency and accuracy.

However, tasks requiring judgment and creativity, such as court representation and judicial decision-making (trust at only 35\%), involve meaning generation---creatively constructing new understanding frameworks within specific contexts \cite{ref22,ref6}. Here, consistency becomes a negative factor that limits innovation and individualized judgment. LLM systems can only reproduce existing bias patterns from training data; they cannot generate transcendent justice understanding in new judicial contexts \cite{ref25}.

Second, an ontological gap exists: the categorical leap from computational rationality to practical wisdom. The root of acceptance differentiation lies in a fundamental gap between two forms of rationality required in judicial practice. Technical tasks can be effectively processed by LLMs, such as e-discovery (despite only 12\% corporate usage, courts widely accept it) and transcription services (22.1\% law firm usage) \cite{ref19,ref20}. These tasks succeed because they can be reduced to computational rationality---optimization processes based on clear rules and quantifiable goals.

In contrast, judicial decision-making requires Aristotelian practical wisdom---the ability to balance multiple values and make prudent judgments in specific contexts \cite{ref26}. When legal consultation is considered as involving professional core values and court representation is viewed as requiring irreplaceable interpersonal trust \cite{ref7}, consistency's limitations in the realm of practical wisdom become evident. This type of wisdom requires emotional resonance, moral intuition, and cultural understanding---embodied and embedded cognitive abilities that mechanical consistency cannot provide \cite{ref27}. Concerns about systematic risk of algorithmic bias in judicial decision-making essentially represent an intuitive rejection of replacing practical wisdom with computational rationality.

Third, an axiological paradox emerges: the challenge of legitimacy reconstruction from procedural propriety to substantive justice. Task acceptance differentiation reflects inherent tensions in judicial legitimacy foundations. High acceptance of technical tasks builds upon procedural legitimacy logic---as long as LLM systems follow established procedures and produce verifiable results, their application possesses formal legitimacy.

However, ultimate judicial legitimacy derives from achieving substantive justice. This requires not only correct procedures but also just outcomes, meaningful processes, and socially recognized judgments \cite{ref28}. The justice threat paradox facing judicial decision support reveals a troubling pattern: the more strictly LLM systems follow formal consistency, the more likely they deviate from substantive justice requirements, thereby undermining their social legitimacy foundation. High hallucination rates in sentencing recommendations (professional legal LLM tools showing 17\%+, Westlaw AI exceeding 34\%) \cite{ref23} represent not merely technical defects but reflect the fundamental contradiction between formalized judicial procedures and the pursuit of substantive justice \cite{ref29}.

The three mechanisms above reveal fundamental flaws in current LLM judicial applications: applying a single instrumental rationality logic to all judicial tasks while ignoring different task types' differentiated consistency requirements. Current LLMs adopt a unified probabilistic calculation model for all task types. This rationality monism strategy inevitably leads to systematic failures in value-intensive tasks. The real solution lies in recognizing consistency's dual-edged characteristics and constructing differentiated task processing frameworks accordingly. This means allowing instrumental rationality to fully leverage consistency's efficiency advantages within its effective domain, while protecting value rationality's dominant position in judicial core functions. Ultimately, through rationality pluralism institutional design, it is possible to resolve consistency's dual-edged effects and create conditions for building social legitimacy foundations.

\subsection{Stakeholder Dimension: Cross-Group Value Rationality Differentiation and Communication Possibilities}

Stakeholder attitudes toward LLM judicial applications present systematic differentiation patterns. This profoundly reflects how consistency's dual nature triggers varied responses among different groups, affecting the establishment of social legitimacy foundations. To comprehensively understand this phenomenon, this study integrated empirical data spanning all levels of the judicial ecosystem from 2023-2025. Table~\ref{tab:stakeholder-analysis} compiles attitude surveys and usage data covering 20 stakeholder groups. Data sources include Thomson Reuters Institute annual judicial reports, American Bar Association technology surveys, Pew Research Center national opinion polls, and behavioral science research from academic institutions including Stanford and Harvard.

\begin{table*}[bt!]
\centering
\small % Reduce font size
\caption{Value Recognition Differences in LLM Judicial Applications from Cross-Group Perspectives}
\label{tab:stakeholder-analysis}
\begin{adjustbox}{width=\textwidth,center}
\begin{tabular}{p{2.8cm}p{3.8cm}p{3.8cm}p{3.2cm}p{1.4cm}}
\toprule
\textbf{Group Type} & \textbf{Main Content} & \textbf{Key Findings} & \textbf{Main Concerns} & \textbf{Research Sources}\\
\midrule
\multicolumn{5}{l}{\textit{Decision Makers}}\\
Judges (USA) & Survey of 1700+ professionals on generative AI attitudes & 9\% believe should use generative LLMs; 68\% express uncertainty & Dehumanization of justice; judicial independence & \cite{ref7,ref30}\\
\addlinespace[3pt]
Judges (Shenzhen) & First judicial vertical LLM deployment covering 85 process nodes & Full system deployment for almost all civil and commercial cases & Over-reliance concerns; responsibility attribution unclear & \cite{ref31,ref2}\\
\addlinespace[3pt]
Judges (UK Welsh) & Judicial Attitude Survey with 94\% salaried judges participating & Open but cautious attitude toward LLM assistance & Manipulation risks; need for clear usage guidelines & \cite{ref32}\\
\addlinespace[3pt]
Lawyers (International) & Survey of 440+ lawyers from large and medium firms on ChatGPT & 82\% believe applicable, but only 3\% actually using generative LLMs & Accuracy concerns (62\%); data security and client privacy & \cite{ref15}\\
\addlinespace[3pt]
Law Firms & Research on firm-level AI policy development and management & 15\% of firms warn against unauthorized use; only 6-41\% have LLM policies & Efficiency pressure vs. replacement fears; need clear policies & \cite{ref15,ref33}\\
\addlinespace[3pt]
Lawyers (USA) & Analysis based on tens of thousands of user data and 1000+ responses & 79\% use LLMs (increased from 19\% in 2023); 74\% hourly work automatable & Concern about impact on revenue model (hourly billing) & \cite{ref14}\\
\addlinespace[3pt]
Lawyers (Florida) & Florida Bar member opinion survey on generative AI & 80\% don't use generative LLMs; 79\% believe "very strict regulation" needed & Cannot replace critical thinking and client explanation ability & \cite{ref34}\\
\addlinespace[3pt]
Legal Professionals & Multiple industry surveys analyzing legal professionals' AI acceptance & 54\% list security issues as major implementation barrier & Job replacement fears; balancing benefits with ethical concerns & \cite{ref35}\\
\addlinespace[5pt]
\multicolumn{5}{l}{\textit{Public and Court Users}}\\
Public (USA) & National survey of 10,701 adults on ChatGPT trust and usage & Only 2\% highly trust ChatGPT for election info; 52\% more worried than excited & Lack of empathy and emotional understanding & \cite{ref36}\\
\addlinespace[3pt]
Court User Public & NCSC \& Thomson Reuters research on AI opportunities in courts & Over 70\% support auxiliary functions; only 27\% believe courts provide adequate help & Functional acceptance vs. decisional resistance contradiction & \cite{ref37}\\
\addlinespace[3pt]
General Public (Trust) & Study on public trust in AI-assisted judges by race/ethnicity & Pure expertise judges most trusted; AI-assisted judges less trusted & AI seen as amplifying existing biases; procedural justice concerns & \cite{ref6}\\
\addlinespace[5pt]
\multicolumn{5}{l}{\textit{Vulnerable Groups}}\\
Minority Groups & AI impact on minority communities' civil rights & Enforcement agencies' AI decisions disproportionately impact immigrants/minorities & AI perpetuating historical biases; lack of transparency & \cite{ref38}\\
\addlinespace[3pt]
Elderly (Judicial) & First case using AI-generated deceased victim video in court & Low acceptance for AI use in court proceedings & Deepfake technology abuse concerns; loss of human element \\
\addlinespace[3pt]
Low-Income Groups & DoNotPay helped overturn \$4M+ fines but fined for false advertising & AI tools help save millions but quality concerns remain & New inequality if high-quality AI legal services require payment & \cite{ref39}\\
\addlinespace[3pt]
Crime Victims & AI-generated deceased victim video used in sentencing hearing & First use may set legal precedent; victim rights groups cautious & Consent and ethical issues; authenticity questions & \cite{ref40}\\
\addlinespace[5pt]
\multicolumn{5}{l}{\textit{Experts and Special Groups}}\\
Legal Scholars & Research on generative AI legal ethics and practical applications & AI will become core legal service technology; existing ethical rules sufficient & "Hallucination" risks; over-reliance weakening independent judgment & \cite{ref41}\\
\addlinespace[3pt]
Compliance Officers & Survey on corporate legal departments' AI compliance confidence & Only 40\% highly confident in LLM-related judicial compliance & IP disputes; deepfake risks in litigation; unclear norms & \cite{ref42}\\
\addlinespace[3pt]
Disabled Persons & AI systems need inclusive design to serve all populations & Court LLM tools often lack accessibility design features & Cannot understand and accommodate special needs in proceedings \\
\bottomrule
\end{tabular}
\end{adjustbox}
\end{table*}

Based on the systematic analysis of stakeholder groups across multiple judicial domains, this study finds that the consistency-acceptability divergence in LLM judicial applications stems from three structural constraints. These constraints collectively reveal how consistency's dual nature hinders the establishment of communicative rationality mechanisms, thereby affecting the construction of social legitimacy foundations.

First, hierarchical power distribution produces a power-acceptance inverse effect. Decision-making power levels show strong conservative tendencies, with only 9\% of U.S. judges believing LLMs should be used and 68\% expressing uncertainty \cite{ref7,ref30}. Although the Shenzhen Court fully deployed an intelligent trial system covering 85 process nodes, management still expresses deep concerns \cite{ref31,ref2}. Among UK Welsh judges, 94\% participated in attitude surveys but maintain cautious positions \cite{ref32}.

Professional power levels exhibit huge gaps between cognition and behavior. While 82\% of international lawyers believe LLMs are applicable, only 3\% actually use them \cite{ref15}. U.S. lawyer usage grew from 19\% in 2023 to 79\%, yet 80\% of Florida lawyers still don't use them \cite{ref14,ref34}. Executive power levels show significantly lagging institutionalized responses, with 15\% of law firms issuing warnings against unauthorized use and only 10\%-41\% establishing LLM usage policies \cite{ref15,ref14}.

Audience power levels demonstrate separation between expectations and trust. While 70\% of the public supports auxiliary functions, only 2\% highly trust LLM key decisions \cite{ref36,ref37}. This gradient difference embodies the power protection mechanism explained by Bourdieu's field theory---groups with higher power positions increasingly view consistency as erosion of their professional autonomy. They tend to maintain existing power structure stability through institutional resistance strategies \cite{ref43}.

Second, uneven distribution of cultural capital forms a professional knowledge-technical concern positive correlation pattern. Legal professionals universally show high concern levels: 54\% list security issues as major barriers \cite{ref35}, 79\% believe in very strict regulation needs \cite{ref34}, legal scholars insist on the centrality of human judgment \cite{ref41}, and only 40\% of corporate compliance officers feel highly confident in LLM judicial compliance \cite{ref42}.

In contrast, the general public presents contradictory attitudes, with 52\% more worried than excited. Special groups like people with disabilities, low-income groups, and the elderly face more specific difficulties \cite{ref36,ref39}. This difference embodies the structural tension between professional rationality and technical rationality within Weber's rationalization theory framework. Professional groups, drawing on their deep legal knowledge and accumulated experience, can more keenly identify consistency's inherent limitations in complex legal reasoning, value judgments, and ethical considerations. This leads them to generate technical caution based on professional responsibility \cite{ref44}.

Third, cultural solidification of value habitus leads to cross-regional culture-technology adaptability differentiation. China demonstrates ambition for large-scale technical practice, with the Shenzhen Court becoming the first to integrate LLMs into judicial decision-making on a large scale, covering almost all civil and commercial cases \cite{ref31,ref2}. The United States presents internal differentiation patterns, with 58\% of judges uncertain about usage \cite{ref7}, while 74\% of lawyers' hourly billing work faces automation threats \cite{ref14}. The UK maintains a prudent attitude, with judges holding open but cautious positions toward LLMs in trial assistance \cite{ref32}.

These regional differences reflect varying understandings of consistency value under different judicial traditions and institutional environments. China's collective interest priority and efficiency-oriented culture views consistency as positive value. America's individual rights protection tradition views consistency as a potential threat. Britain's incremental reform tradition seeks balance between the two \cite{ref2}.

The interaction of these three constraints leads to the absence of communicative rationality mechanisms, preventing consistency's technical advantages from translating into social recognition. Instead, they become obstacles to establishing legitimacy foundations. This manifests in three aspects: institutional supply-demand imbalance between regulatory needs and policy provision, with 60\% completely unaware of LLM detection tools \cite{ref34}; cognitive-behavioral deviation between acceptance and practice, with 51\% believing in application but only 3\% actually using \cite{ref15}; and attitude differentiation based on field positions rather than common value pursuits. Crime victims worry about LLM evidence authenticity \cite{ref40}, while disabled persons face technological exclusion.

Due to the lack of situations emphasizing equal participation, mutual understanding, and consensus orientation as outlined in communicative rationality theory, stakeholders respond strategically based on partial understandings of consistency rather than seeking communicative understanding based on common interests and value consensus. This prevents effective mediation of the fundamental tension between formal rational consistency pursued by technical systems and value rational pluralism upheld by social groups through democratic deliberation. Ultimately, this leads to the solidification and institutional reproduction of the consistency-acceptability divergence phenomenon.

\section{Discussion}

\begin{figure*}[bt!]
\centering
\includegraphics[width=0.8\textwidth]{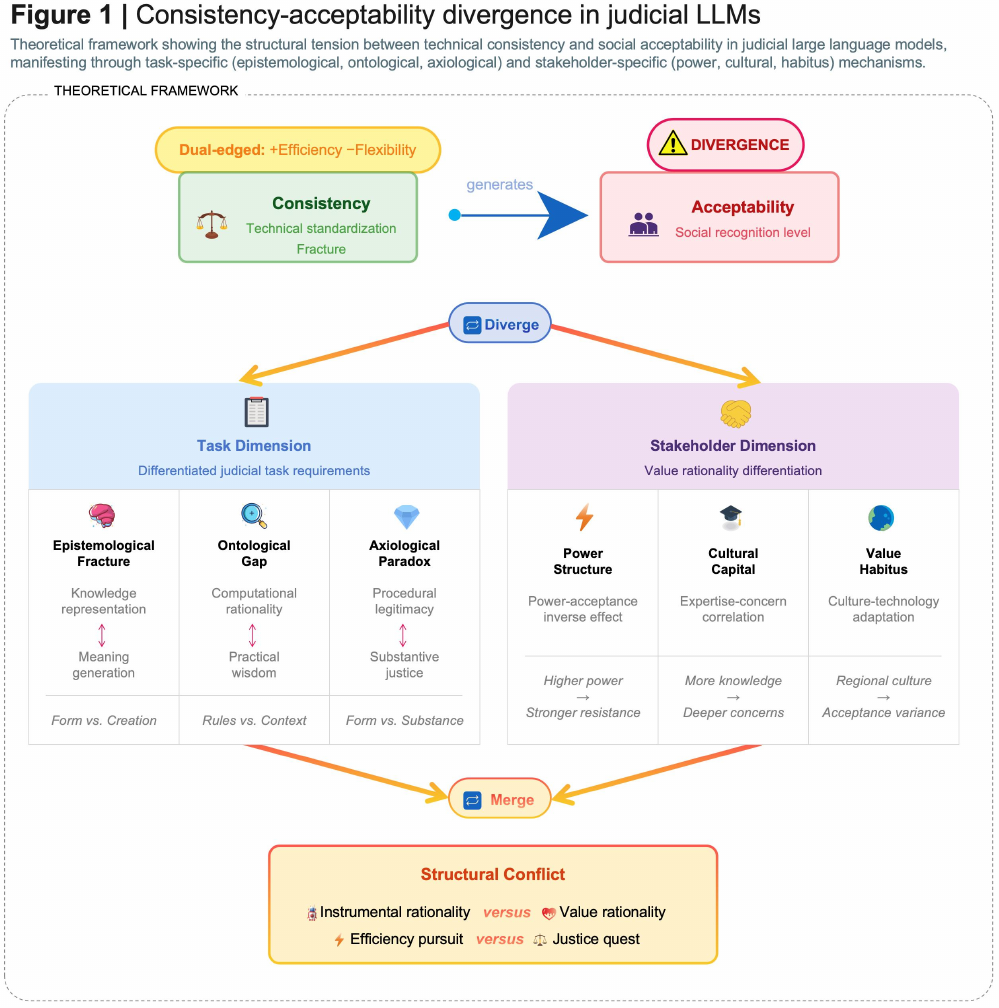}
\caption{Consistency-acceptability divergence in judicial LLMs. The theoretical framework showing the structural tension between technical consistency and social acceptability in judicial large language models, manifesting through task-specific (epistemological, ontological, axiological) and stakeholder-specific (power, cultural, habitus) mechanisms.}
\label{fig:theoretical-framework}
\end{figure*}

Through systematic analysis, this study reveals the core paradox of consistency-acceptability divergence in LLM judicial decision-making. This study finds that this divergence stems from structural tensions between consistency's dual characteristics and value rationality requirements. As illustrated in Figure~\ref{fig:theoretical-framework}, this divergence manifests through two key dimensions: in the task dimension, it appears as three mechanisms---epistemological fracture, ontological gap, and axiological paradox; in the stakeholder dimension, it emerges as three constraints---power-acceptance inverse effect, professional knowledge-technical concern positive correlation, and culture-technology adaptability differentiation \cite{ref6,ref45}. The theoretical framework presented in Figure~\ref{fig:theoretical-framework} demonstrates how these dual dimensions converge to create the fundamental structural conflict between instrumental rationality (efficiency pursuit) and value rationality (justice quest). Theoretically, proposing the consistency-acceptability divergence conceptual framework holds innovative significance. The analytical perspective of this study integrating Weber's rationalization theory and Habermas's communicative rationality theory reveals the possibility of achieving dynamic balance between instrumental and value rationality through communicative approaches. Practically, the findings of this study provide differentiated AI application strategy guidance for judicial institutions. For technical tasks, institutions should fully leverage consistency's efficiency advantages. For value judgment tasks, they must establish strict human supervision mechanisms. Through the DTDMR-LJGF framework, institutions can construct multi-stakeholder deliberation mechanisms. This research also warns of risks from the marginalized group paradox and forced adoption phenomena, providing direct guidance for formulating balanced, prudent, human-centered judicial AI policies \cite{ref46}.

\begin{figure*}[bt!]
\centering
\includegraphics[width=0.7\textwidth]{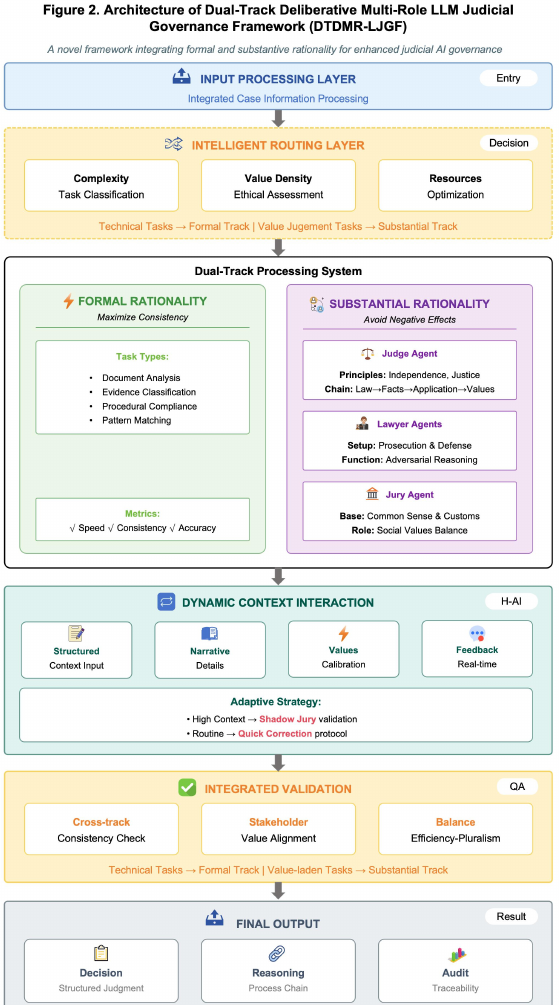}
\caption{Architecture of Dual-Track Deliberative Multi-Role LLM Judicial Governance Framework (DTDMR-LJGF). A novel framework integrating formal and substantial rationality for enhanced judicial AI governance.}
\label{fig:dtdmr-framework}
\end{figure*}

To address the consistency-acceptability divergence, this study proposes the Dual-Track Deliberative Multi-Role LLM Judicial Governance Framework (DTDMR-LJGF). This framework fundamentally reshapes LLM judicial governance logic. As shown in Figure~\ref{fig:dtdmr-framework}, it achieves precise task routing through an intelligent routing layer. Procedural tasks are directed to the formal rationality track for efficient processing, while value judgment tasks activate the substantive rationality track's multi-role deliberation mechanism (including judge agents, lawyer agents, and jury agents). The framework innovatively introduces a dynamic context interaction interface as a bidirectional interactive value calibration space for deep human-machine integration. Relying on a three-layer dynamic architecture---input processing layer, dual-track processing system, and integrated validation system---it maintains consistency advantages while preserving value pluralism. Through differentiated processing strategies using shadow jury mechanisms and rapid correction mechanisms, it achieves the practical implementation of communicative rationality within technical systems. This provides an operational solution that leverages technical advantages while maintaining judicial values for establishing social legitimacy foundations \cite{ref47}.

In terms of research perspective and solutions, this paper achieves important breakthroughs. By taking consistency's dual characteristics as the core analytical concept and constructing a task-stakeholder dual-dimensional analytical framework, this study achieves for the first time a three-dimensional holistic understanding of LLM judicial applications. This breaks through the fragmented landscape of technical performance evaluation, social acceptance research, and ethical analysis \cite{ref25}.

Through systematic integration of multi-source data from 2023-2025, this study constructs the most comprehensive empirical dataset of LLM judicial applications to date. Cross-validation reveals differentiated effect patterns of consistency across different contexts \cite{ref2}. Theoretically, by tracing surface phenomena to deep tensions between instrumental and value rationality, the DTDMR-LJGF framework represents a fundamental reconstruction of AI system architecture based on communicative rationality theory. By embedding multi-stakeholder deliberation processes that incorporate diverse values into AI operational mechanisms, it indicates a technically feasible and ethically sound direction for judicial AI development.

However, this study has several limitations. First, there is geographical and cultural imbalance in data collection, with samples mainly concentrated in developed regions of North America, Europe, and East Asia. Insufficient attention to developing regions may limit the full capture of consistency's differentiated manifestations across cultural contexts. Second, the temporal window concentrated in 2023-2025 makes it difficult to observe long-term evolution patterns. Third, methodological constraints from original research limitations mean successful cases are more easily cited while failure experiences often remain hidden. Fourth, theoretically, translating communicative rationality from philosophical theory into specific technical design and institutional arrangements requires further exploration. These limitations clarify the research's scope while indicating directions for future research: expanding geographical coverage, conducting long-term tracking, establishing standardized evaluation systems, and advancing theory-practice translation.

\section{Materials and Methods}

This study comprehensively employed interdisciplinary perspectives and methods crossing legal philosophy and artificial intelligence computational science. Under this framework, this study proposed the theoretical innovation of consistency-acceptability. Building on this foundation, this study adopted systematic literature analysis to examine empirical research and practical data on LLM judicial application acceptance published between 2023 and 2025 \cite{ref48}.

Literature searches covered major academic databases including Web of Science, Scopus, Google Scholar, and SSRN. This study used combinations of keywords such as ``large language models,'' ``LLMs,'' ``ChatGPT,'' ``AI'' with ``judicial,'' ``legal,'' ``court,'' ``justice,'' ``acceptance,'' ``adoption,'' and ``perception'' \cite{ref49}.

After systematic screening \cite{ref50}, the final dataset integrated four main source types: annual survey reports from industry research institutions (Thomson Reuters Institute, Clio, Bloomberg Law, American Bar Association, etc.); empirical research from academic institutions (Stanford RegLab, Harvard Law School, University of Pennsylvania, etc.); judicial practice data (Shenzhen Intermediate Court, U.S. Federal Courts, UK Welsh Courts, etc.); and public opinion surveys (Gallup, Pew Research Center, National Center for State Courts, etc.). Data collection methods included questionnaire surveys, judicial case analysis, controlled experiments, and in-depth interviews. These multi-source data collectively constitute the empirical foundation for analyzing LLM judicial application acceptance.

\section*{Acknowledgments}

We thank the School of Software and School of Law at Yunnan University for providing research facilities and computational resources. We are grateful to the anonymous reviewers for their constructive comments that improved this manuscript.

\textbf{Funding:} This research received no specific grant from any funding agency in the public, commercial, or not-for-profit sectors.

\textbf{Author contributions:} M.Z. conceived the research framework, conducted the systematic literature analysis, developed the theoretical framework, and wrote the manuscript. Q.X. contributed to conceptual development, provided legal expertise, critically revised the manuscript, and approved the final version.

\textbf{Competing interests:} The authors declare no competing interests.

\textbf{Data availability:} All data analyzed are from publicly available sources cited in the references.

%Bibliography
%\bibliographystyle{unsrt}  
%\bibliography{references}  

% References section

\end{document}